\documentstyle[aps,preprint,epsfig]{revtex}
\newcommand{\r}{\rho}

\begin{document}
\begin{titlepage}
\begin{centering}
\title{Next to Leading Order Semi-inclusive Spin Asymmetries\thanks{Partially 
supported by CONICET-Argentina.}}
\author{{ D.de Florian, L.N.Epele, H. Fanchiotti, C.A.Garc\'{\i}a Canal}}
\address{
Laboratorio de F\'{\i}sica Te\'{o}rica,
Departamento de F\'{\i}sica \\
Universidad Nacional de La Plata,  C.C. 67 - 1900 La Plata - 
Argentina}
\author{
 { S.Joffily}}
\address{ 
 Centro Brasilero de Pesquisas Fisicas \\ Rua Xavier Sigaud 150, Urca, 22.290.180 Rio de Janeiro - Brazil}
\author{  { R.Sassot} }
\address{Departamento de F\'{\i}sica, 
Universidad de Buenos Aires \\ 
Ciudad Universitaria, Pab.1 
(1428) Bs.As. - 
Argentina}
\date{15 March 1996}
\maketitle
\end{centering}

\begin{abstract}
We have computed semi-inclusive spin asymmetries for proton and deuteron targets including
next to leading order (NLO) QCD corrections and contributions coming from the target fragmentation region. These corrections  have been estimated using NLO fragmentation functions, parton distributions and also a model for spin dependent fracture functions which is proposed here.
We have found that NLO corrections are small but non-negligible in a scheme where gluons 
are polarised and that our estimate for target fragmentation effects does not modify significantly charged asymmetries but affects the so called difference asymmetries.

\end{abstract}

\end{titlepage}

\noindent{\large \bf Introduction:}\\

Recently, the Spin Muon Collaboration (SMC) \cite{smc} have presented a measurement of semi-inclusive
spin asymmetries for positively and negatively charged hadrons from deep inelastic scattering 
of polarised muons on polarised protons and deuterons. This data, combined with previous measurements \cite{emc,wisl,baum} of this kind  can be used to determine polarised 
valence and non-strange sea quark distributions, independently from totally inclusive data.

Up to now, the analyses \cite{smc,emc,wisl,baum} of semi-inclusive spin asymmetries have been performed in the naive 
quark-parton model, neglecting both higher order corrections and contributions coming 
from the target fragmentation region. This procedure simplifies greatly the extraction of 
parton distributions and seems to be adequate given the present accuracy of the data and the
restriction to high hadron energy fractions.

However, taking into account that the most recent analyses of parton distributions \cite{nosprd,vogelsang,stirling}, which are 
performed in the NLO approximation from totally inclusive data, have shown the importance of including these effects, and that the forthcoming semi-inclusive experiments \cite{proposal} promise better accuracy than the obtained so far, it is worthwhile analysing the size of these hitherto neglected contributions.

Higher order corrections can be non-negligible if gluons are polarised in the proton and, in the case of
semi-inclusive processes, require a non trivial treatment of collinear divergences related to the target fragmentation region. This has been addressed in references \cite{altarelli} and \cite{veneziano}. In this last reference,  
the concept of fracture functions has been introduced as a mean to describe target fragmentation phenomena. The full NLO contributions to semi-inclusive cross-sections, including those related to fracture functions, have been calculated recently in references \cite{graudenz} and \cite{nosnpb} for unpolarised and polarised deep inelastic scattering, respectively.

As the parton distributions and the fragmentation functions, fracture functions are non-pertubative 
objects that have to be extracted from semi-inclusive high precision experiments. This
task is not possible yet, however one can estimate the size of the target fragmentation corrections effects using a sensible model for fracture functions
based on parton model ideas.

In order to establish our notation, in the next section we show the naive quark parton model expressions for the semi-inclusive cross sections and the full NLO ones. We also remind
the definition of the spin asymmetries in terms of the former cross sections. In the following
section we present our choice for parton distributions, fragmentation functions and the main
features of the model proposed for fracture functions. In the last section we show results
and present our conclusions.\\

\noindent{\large \bf NLO Cross Sections}\\

In the naive quark-parton model, the semi-inclusive cross section for the production of a hadron $h$ from polarised deep inelastic scattering of charged leptons carring momentum $l$ on nucleons with momentum $P$, is usually written as \cite{nosnpb}:
\begin{eqnarray}
 \frac{d\Delta \sigma^h_N}{dx\,dy\,dz} & = &
  \lambda Y_P\sum_{i=q,\bar q} c_i\, \Delta q_i(x)\, D_{h/i}(z)
\end{eqnarray}
where $\lambda$ is the helicity of the lepton, $c_{i}=4\pi e_{q_{i}}^2 \alpha^2/x(P+l)^2$, and
$\Delta \sigma^h_N$ denotes the difference between cross sections of targets with opposite helicities.  This cross section is differential in the variables $x$, $y$ and $z$ defined by
\begin{equation}
x=\frac{Q^2}{2 P\cdot q} , \ \ \ y=\frac{P\cdot q}{P\cdot l} ,\ \ \ z=\frac{E_h }{E_N (1-x)}
\end{equation}
where $q$ is the transfered momentum $(-q^2=Q^2)$ and $E_h$ and $E_N$ are 
the produced hadron and target nucleon energies, respectively.
The unpolarised cross section can easily be obtained changing the kinematical factor $\lambda Y_P=\lambda (2-y)/y$ for $Y_M=(1+(1-y)^2)/2y^2$ and removing the $\Delta$'s, which denote differences in polarization. $\Delta q_{i}$ is the spin-dependent parton distribution of flavour $i$ and 
$D_{h/i}$ is the fragmentation function of a hadron $h$ from a parton $i$.

It is customary to define spin asymmetries $A_{1\,N}^{h}$, proportional to the difference between the number of events for antiparallel and parallel orientation of the lepton and the 
nucleon spins, which in our notation are given by
\begin{equation}
A_{1\,N}^{h}=\frac{Y_M}{\lambda Y_P}\frac{\Delta \sigma^h_{N}}{\sigma^h_{N}}
\end{equation}
and in the naive parton model reduce to
\begin{equation}
A_{1\,N}^{h}=\frac{\sum_{i} e^2_{i}\Delta q(x) D_{h/i}(z)}{\sum_{i} e^2_{i} q(x) D_{h/i}(z)}
\end{equation}
Actually, the data on these asymmetries is restricted to positively and negatively charged hadrons  with the cross section integrated over some range of the variable $z$.

The difference asymetries \cite{frankfurt}, $A_{N}^{h^+-h^-}$ are given by 
\begin{equation}
A_{N}^{h^+-h^-}=\frac{Y_M}{\lambda Y_P}\frac{\Delta \sigma^{h^+}_{N}-\Delta \sigma^{h^-}_{N}}
{\sigma^{h^+}_{N}-\sigma^{h^-}_{N}}
\end{equation}
and in this approximation have no dependence on the fragmentation functions, leading to expressions
like
\begin{equation}
A_{D}^{\pi^+-\pi^-}=\frac{\Delta u_v +\Delta d_v}{u_v+d_v}, \,\,\,\,\,\,\,\,\,\,\,\
A_{p}^{\pi^+-\pi^-}=\frac{4 \Delta u_v -\Delta d_v}{4 u_v-d_v}
\end{equation}
for pion production on deuterium and proton targets respectively.
In the next to leading order approximation, the polarised cross sections have the following expression
\begin{eqnarray}
 \frac{d\Delta \sigma^h_N}{dx\,dy\,dz}  =
  \lambda  _P \sum_{i=q,\bar q} c_i \left\{\int \int_{A} \frac{du}{u}
\frac{d\r}{\r} 
\right. 
 && \left. \left\{ \,
\Delta q_i(\frac{x}{u},Q^2)\, D_{h/i}(\frac{z}{\r},Q^2) \,
\delta(1-u)\delta(1-\r)\frac{}{} 
\right. \right.  \nonumber \\
 && +   \left.\left.
\Delta q_i(\frac{x}{u},Q^2)\, D_{h/i}(\frac{z}{\r},Q^2)  \,
 \Delta C_{qq}(u,\r)   \right. \right. \nonumber \\
 && + \left. \left. \Delta q_i(\frac{x}{u},Q^2)\, D_{h/g}(\frac{z}{\r},Q^2)\,
 \Delta C_{qg}(u,\r)  
 \right. \right.\nonumber\\
&& +  \left. \left. \Delta g(\frac{x}{u},Q^2)\, D_{h/i}(\frac{z}{\r},Q^2)\,
  \Delta C_{gq}(u,\r) 
\right\}  \right. \nonumber\\ 
 + \left.  \int_{B} \frac{du}{u} (1-x)
\right. && \left. \left\{ \,  \Delta M^h_{q_i}(\frac{x}{u},(1-x) z,Q^2)  \left(
\delta(1-u) \frac{}{} + 
 \Delta C_{q}(u)  \right)
 \right.\right. \nonumber \\ 
 &&  +  \left.\left. \Delta M^h_g (\frac{x}{u},(1-x) z,Q^2) \,
\Delta C_g(u)
\right\}\right\} 
\end{eqnarray}
where the $\Delta C$'s are the NLO coefficient functions \cite{coef}, which are proportional to $\alpha_s$, and $\Delta M^h_i$ are the spin dependent fracture functions. Details about the convolution variables and integration limits can be found in references \cite{graudenz,nosnpb}.
Notice that the difference between equations (1) and (7) is not only proportional to $\alpha_s$, but there is a leading order fracture contribution which is neglected in the most naive approximation.
Obviously, the spin asymmetries develop much more complicated expressions, particularly, the difference asymmetries do not reduce just to combinations of partons distributions as in equation (6), and depend on the variable $z$.
Notice that the variable $z$ defined in equation (2) and used in equation (7) coincides with the one used in the analyses performed up to now, $z_h= P.P_h/P.q$ \cite{altarelli} in the naive approximation but they differ for higher order processes, in which the hadron may be produced at an arbitrary angle $\theta$ with respect to the beam direction
\begin{equation}
z_h = z\, \frac{1+\cos \theta}{2}
\end{equation}
The $z$ variable so defined  is much more convenient for factorization purposes \cite{graudenz}.\\

\noindent{\large \bf Inputs}\\

In order to feed equation (7) with parton distributions and fragmentation functions, we have chosen two sets of NLO parametrizations for polarised parton distributions \cite{nosprd51}, one for unpolarised distributions \cite{mrs} and one for NLO fragmentation functions \cite{kramer}.
The polarised sets reproduce the main features of the available inclusive data and are defined within a physical factorization presciption ($\overline{MS_p}$), the same chosen for the coefficients in equation (7). In one of these sets (set 1) the gluons are polarised whereas in the other (set 2), the strange sea quarks are responsible for the low value of Ellis-Jaffe integral \cite{ellis}.
Both sets satisfy positivity constraints with respect to the unpolarised sets, something that is crucial for computing asymmetries.
The fragmentation functions do not imply the full flavour symmetry relations between hadrons that were assumed in reference \cite{smc}. These functions were obtained as  independent NLO fits to charged pion and kaon production in $e^+ e^-$ annihilation. 

Fracture functions are a relatively new concept and have not been measured yet, so there are not parametrisations available for them.
However, taking into account that these functions measure the probability for finding a hadron and a struck parton in a target
nucleon, one can  approximate them as  a simple convolution products between known distributions. These are the probabilities for finding the struck parton in a nucleon carrying a fraction $x$ of its momentum, the one for finding another parton in the target remnant (with momentum fraction constrained to the interval $[0,1-x]$) and that for its fragmentation in the observed hadron with momentum fraction $z (1-x)$. 
Assuming that the  correlation between both subprocesses is dominated by the momentum balance, a tipically partonic assumption, our proposal for the fracture functions reads as
\begin{equation}
M^h_j (x, z (1-x) )= q_j(x)\, \frac{1}{N(x)} \int_z^1\frac{dt}{t}\sum_k q_k(t (1-x)) D_{h/k} (z/t)
\end{equation}
The index $j$ reffers to the struck parton  (quark or gluon), and $k$ denotes an intermediate parton which undergoes hadronization into a particle $h$. A sum over all possible intermediate flavors and momentum fractions is implied. 
The function $N(x)$, given by
\begin{equation}
N(x)=\frac{\int_0^{1-x}dy\, y\, q(y)}{(1-x)},
\end{equation}
 normalizes the full remnant momentum to $(1-x)$, as required for consistency, and
guarantees the momentum sum rule fulfilment \cite{veneziano}
\begin{equation}
\sum_h \int dz\,z\,M^h_j(x,z(1-x))=\frac{q_j(x)}{(1-x)}
\end{equation}
provided 
\begin{equation}
\sum_h \int dz\,z\,D_{h/j}(z)=1
\end{equation}

Analogously, spin dependent fracture functions can be modelized using spin dependent parton distributions for the struck parton and unpolarized distributions for the remaining part. The normalization function is the same as in  equation (10), which also guarantees the analogous sum rule
\begin{equation}
\sum_h \int dz\,z\,\Delta M^h_j(x,z(1-x))=\frac{\Delta q_j(x)}{(1-x)}
\end{equation}

In the next section we estimate the higher order corrections to the semi-inclusive charged and difference spin asymetries using the distributions presented here and our model for fracture functions.

\noindent{\large \bf Results:}\\

 In order to compare with the
available data on semi-inclusive spin asymmetries, we compute them  taking into account the production
of charged pions and kaons and we integrate the cross sections in the variable $z$ over the measured range. Charged kaon production adds negligible contributions to charged asymmetries, which are dominated by pion production,
however we take them into account because of its role in difference asymmetries as it will be discussed later.

In figure (1) we show positively (1a,1b) and negatively (1c,1d) charged hadron asymmetries on protons using both sets for polarised parton distributions. The solid lines correspond to the most naive contribution -\cal{O}($\alpha_s^0$) and
without target fragmentation effects-, long-dashed lines include NLO corrections
to fragmentation processes, short-dashed lines  (almost overlapping with the solid ones) takes into account fragmentation and fracture but at LO, finally the dotted lines (overlapping with the long dashes) show the result of the full
NLO computation (equation 7). 

These figures show clearly that target fragmentation effects are negligible in the charged
asymmetries for $z>0.2$. This is due, at small $x$ where target fragmentation effects are large,
to the suppression of the full asymmetries caused by the increase of the unpolarised cross section.
At intermediate $x$,  the dominance of current fragmentation
over target fragmentation is the main reason for the smallness of the correction. At larger values of $x$, target fragmentation becomes  again of the same order of current fragmentation, however both hadronization contributions tend to be cancelled in
the asymmetry due to the fact that those considered here -producing spinless final states- are essentially independent of the initial state polarisation (that of the struck parton).
The model accounts for this fact  because it defines fracture functions in which the hadronization part is the same for the polarised and the unpolarised case, being the spin dependence restricted to the probability of finding the struck parton. In other words,
\begin{equation}
\frac{\Delta M^h_i}{M^h_i}=\frac{\Delta q_i }{q_i}
\end{equation}

Next to leading order corrections are small but non negligible for sets with gluon polarization, as can be seen in figures (1a) and (1c), if the forthcoming experiments reach the expected accuracy. As these corrections are dominated by those of gluon origin, they have no significant consequences for set 2, figures (1b) and (1d).  

The same features are observed for deuterium targets, figures (2a), (2b), (2c) and (2d). We also
show the most recent SMC proton and deuterium data \cite{emc}, and that presented by EMC \cite{smc}.

In figures (3a) and (3b) we show the curves of figure (1a) and (1c) but only for
charged pion production against the accuracy expected from Compass \cite{proposal} for two years running at 100 GeV.

It is interesting to notice that the cancellation of fracture function contributions in charged asymmetries allows a naive interpretation for them with less stringent cuts in $z$ than those
used up to now. This choice would eventually allow a substantial improvement of the experimental statistics.
In figure (4) we show the corrections exhibited  in figure (1a) but for $z>0.1$.

A completely different situation is observed for the difference asymetries, figures (5) and (6), in this case calculated for values of $z$ greater than $0.25$ in order to compare with the experimental data presented in reference \cite{baum}. 
In these asymmetries the suppression due to the unpolarised cross sections is not present at small $x$
so target fragmentation effects are then quite significant (short dashes for
LO and dots for NLO) 
This is related to the fact that the asymmetries depend on the differences between the probabilities for positive and negative hadron production. This also  affects the  intermediate $x$ region, where the differences
are comparable in the current and target fragmentation cases.

Corrections to equation (1) have also other serious consequences in difference asymmetries.
Notice that the passage from equation (5) to equation (6) implies the cancellation of a factor, both in the numerator and the denominator of these equations, like
\begin{equation}
\left[ D_{\pi^+/u}(z)-D_{\pi^-/u}(z) \right]
\end{equation}
which is found to be zero around $z\sim 0.2$ in different experiments \cite{kramer}. The above mentioned corrections, however, shift the zeros of numerator and denominator in a different way causing large distortions (even divergencies) from the naive expectation \cite{frankfurt}.  
These distortions make  pointless a naive interpretation of the difference asymmetries, at least for values of $z$ of the order or lower than $0.2$. 
Notice also that in the region where $\pi^+$ production equals that of $\pi^-$, kaon production acquires relevance and can not be neglected

At variance with the charged asymmetries, difference asymmetries only allow analyses with a lower cut in $z$ if large corrections are taken into account.
This fact by no means challenges difference asymmetries. On the contrary, the comparison between the results comming from them and those from inclusive and the other semi-inclusive observables allows one
to explore new aspects of the parton model, in particular the hypotheses of the model presented here.
\\

 \noindent{\large \bf  Conclusions:}\\

In this paper we have found that that NLO corrections to semi-inclusive spin asymmetries, particularly those related to target fragmentation effects, are not negligible and can be
treated quantitatively using a sensible model for fracture functions. 

Taking into account these corrections, one can safely reduce the kinematical cuts used in the
analysis of the experimental data on charged asymmetries, correct the naive interpretation
of the difference asymmetries for $z>0.25$, and give an explanation to the features of the data
for lower $z$ cuts. We expect that these issues will be relevant in the analyses of the
forthcoming semi-inclusive experiments.\\

\noindent{\large \bf Acknowledgements:}\\

We warmly acknowledge R.Piegaia for fruitfull discussions, and F.Perrot Kunne and W.Wi\'slicki for kind help with the experimental data.
\pagebreak

\pagebreak

\noindent{\large \bf Figure Captions:}\\

\begin{itemize}
\item[Figure 1:] Semi-inclusive asymmetries for muoproduction of charged pions and kaons on a proton target with $z>0.2$; a) and b) for positive hadrons calculated with sets 1 and 2, respectively, c) and d) for negative hadrons.
The curves correspond to the naive estimate (solid), adding target fragmentation effects al LO (short dashes almost superimposed with the previous), current fragmentation at NLO (long dashes), and the full NLO prediction. The data correspond to EMC and SMC experiments.
\item[Figure 2:] The same as in figure (1) but for deuterium targets.
\item[Figure 3:] The same as in figures (1a) and (1c) but for charged pion production. The error bars represent the expected accuracy for two years of running of the Compass experiment.
\item[Figure 4:] The same as in figure (1a) but for $z>0.1$
\item[Figure 5:] Naive estimate of the difference asymmetry and corrections, calculated with $z>0.25$ for porton targets.
\item[Figure 6:] The same as in figure 5 but for deuteron targets.   
\end{itemize}

\pagebreak 
\setlength{\unitlength}{1.4mm}
\begin{figure}[hbt]
\begin{picture}(100,110)(7,30)
\mbox{\epsfxsize16.0cm\epsffile{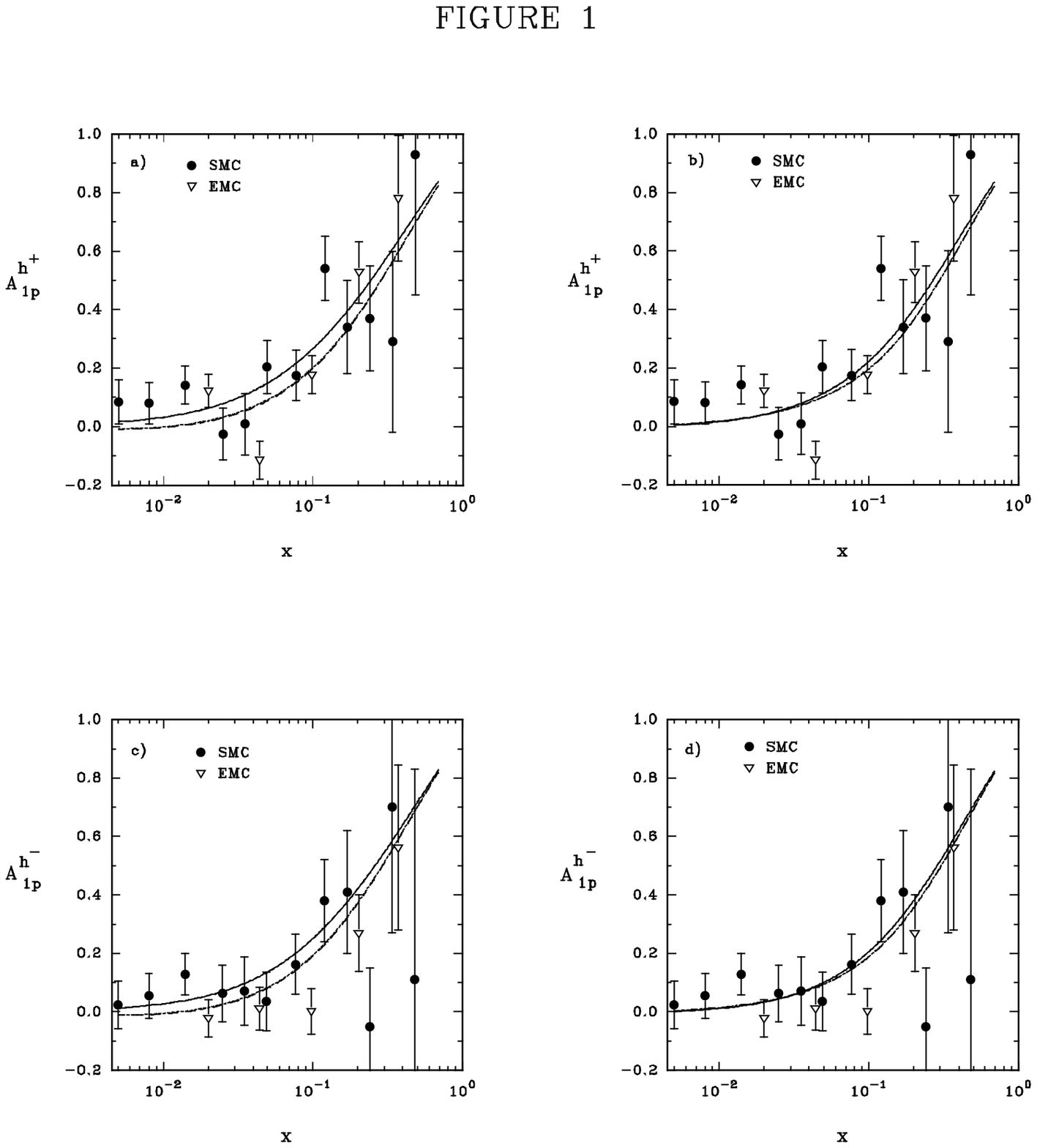}}
\end{picture}
\end{figure}
\pagebreak 
\setlength{\unitlength}{1.4mm}
\begin{figure}[hbt]
\begin{picture}(100,110)(7,30)
\mbox{\epsfxsize16.0cm\epsffile{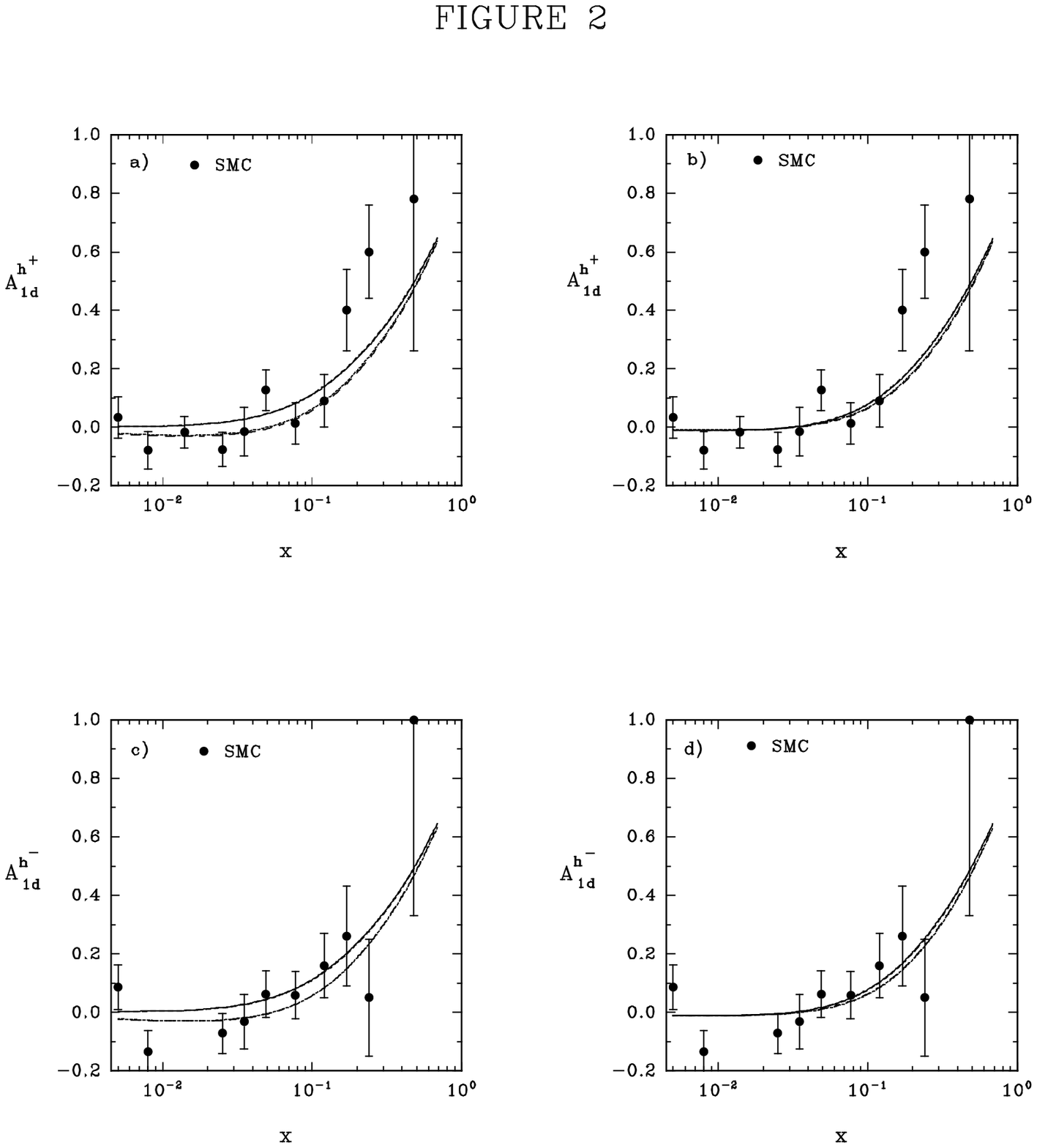}}
\end{picture}
\end{figure}
\pagebreak
\setlength{\unitlength}{1.3mm}
\begin{figure}[hbt]
\begin{picture}(100,110)(2,55)
\mbox{\epsfxsize14.0cm\epsffile{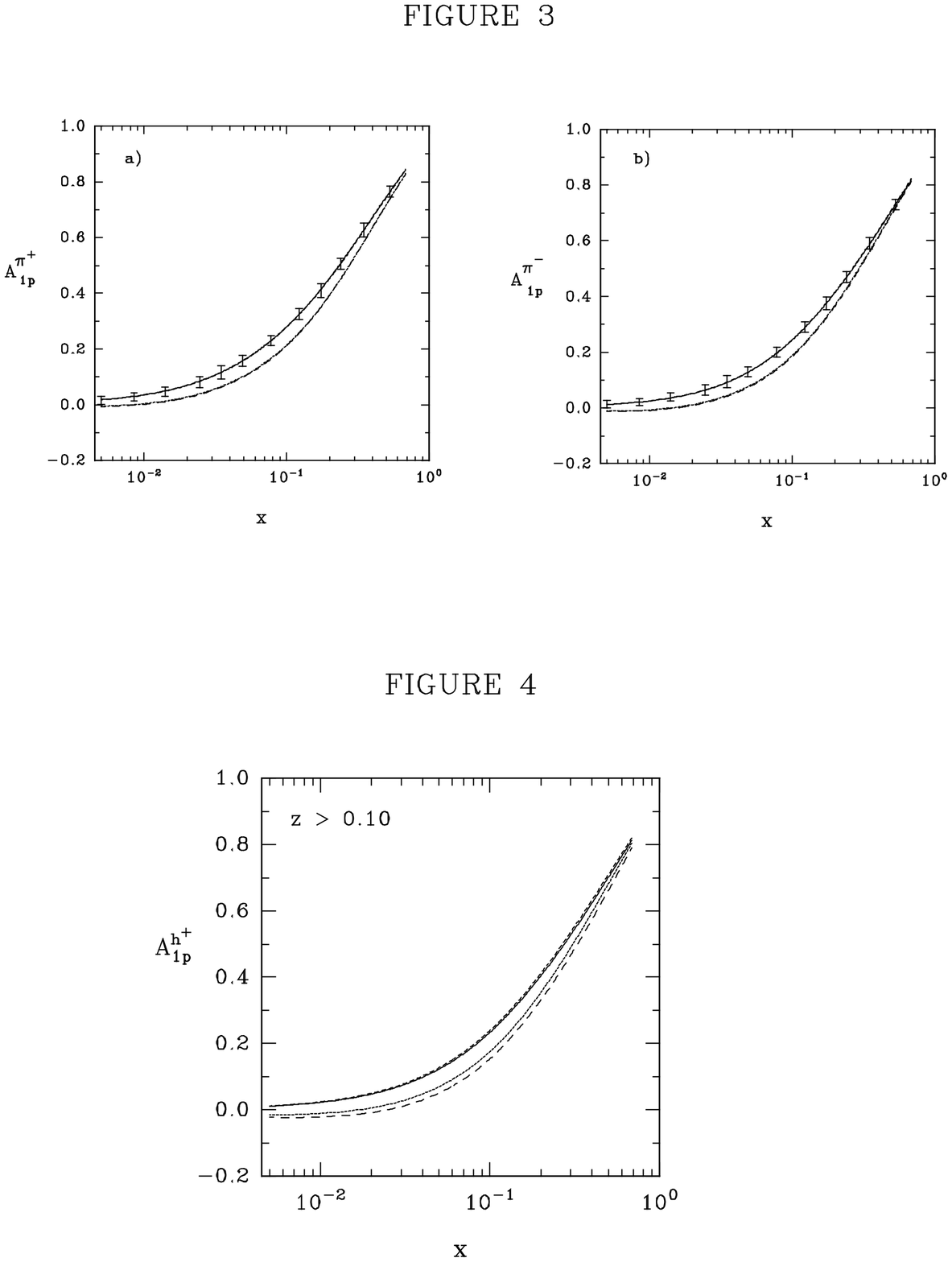}}
\end{picture}
\end{figure}
\pagebreak

\setlength{\unitlength}{1.3mm}
\begin{figure}[hbt]
\begin{picture}(100,110)(2,55)
\mbox{\epsfxsize14.0cm\epsffile{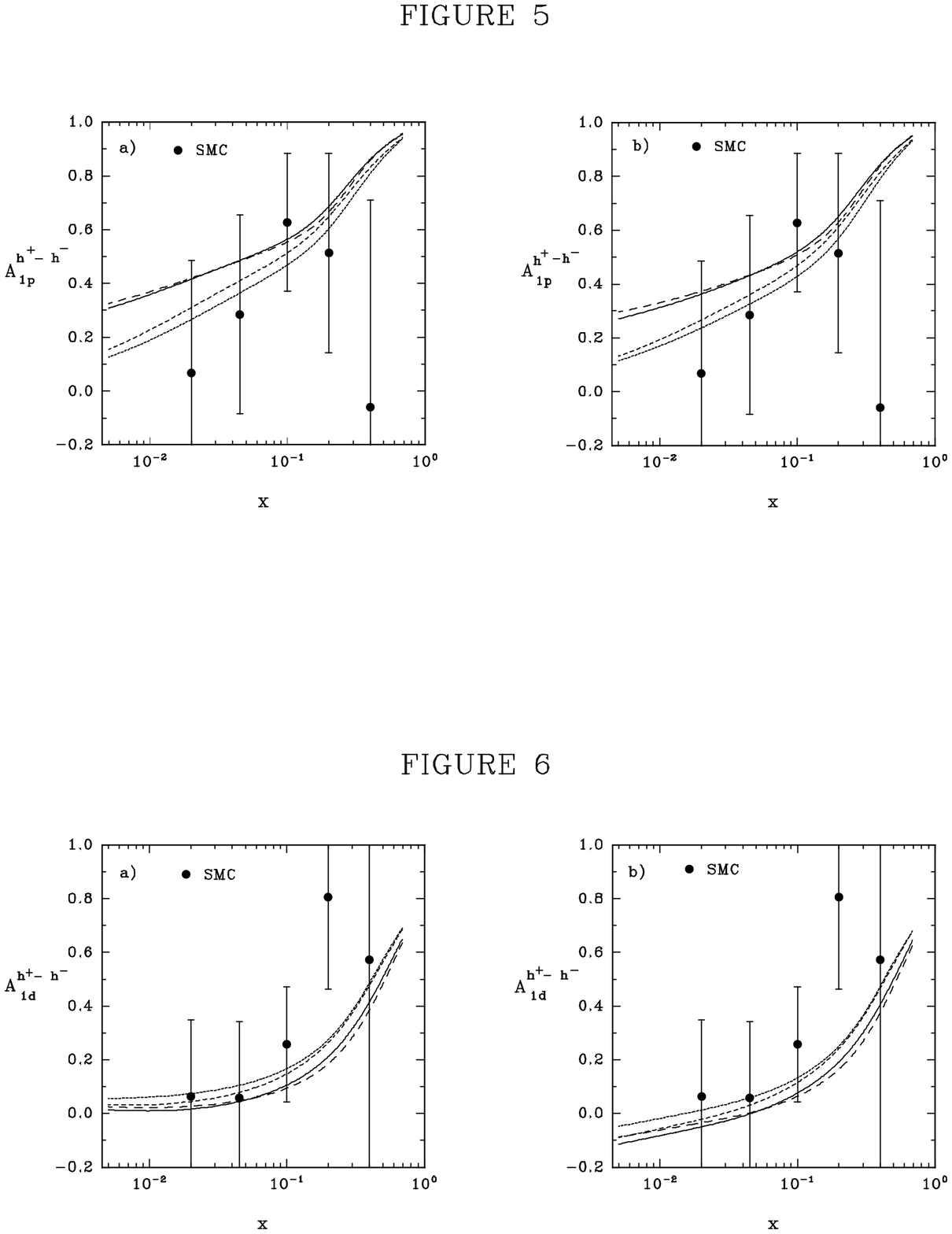}}
\end{picture}
\end{figure}


\begin{references}
\bibitem{coef} These coefficients can be traced back from reference [12].
\end{references}
\end{document}